\documentstyle[aps,pra]{revtex}

\begin{document}


\twocolumn[\hsize\textwidth\columnwidth\hsize\csname
@twocolumnfalse\endcsname

\title{Explicit Euler method  for solving the time dependent Schr\H{o}dinger equation}
\author{Ioan Sturzu\footnote{Email: sturzu@unitbv.ro}}
\address{''Transilvania'' University Brasov, Department of Physics\\
Eroilor 29, Brasov, Romania}
\date{\today}

\maketitle

\begin{abstract}
Using an explicit Euler substitution, a system of
differential equations was obtained, which can be used to find the solution of
the time-dependent 1-dimensional Schr\H{o}dinger equation for a general form of
the time-dependent potential.
\end{abstract}


\vskip1pc]

In the framework of fundamentals of Quantum mechanics, the unitary evolution
given by the Schr\H{o}dinger equation is conceivable only for quasi-isolated
quantum systems, i.e. systems interacting via ''classical'' fields, which
may be obtained using quantification procedures. Actually, these fields are
time-independent, or have a harmonic time-dependence. The evolution of a
quantum system interacting with a non-stationary environment is given by
master-type equations rather than time-dependent Schr\H{o}dinger equations.
However, the latter are often used in many practical applications from
molecular physics, quantum chemistry, quantum optics, solid state physics,
etc. \cite{guedes}.

In the recent paper \cite{guedes}, I. Guedes obtained the exact Schr\H{o}%
dinger wave function for a particle in a time-dependent 1-dimensional linear
potential energy. For a Hamiltonian: 
\begin{equation}
H(x,p,t)=\frac{p^2}{2\cdot m}+f(t)\cdot x  \label{ham}
\end{equation}
if Schr\H{o}dinger equation is tested with the trial function: 
\begin{equation}
\Psi (x,t)=Ne^{[\eta (t)\cdot x+\mu (t)]}  \label{trialG}
\end{equation}
one obtains two differential equations for the functions $\eta (t)$ and $\mu
(t),$ so one can find the solution $\Psi (x,t)$ after solving these
equations. To this end, one has to find the initial conditions, say from $%
\Psi (x,0)$. However, $\Psi (x,0)$ is not, generally, of the type (\ref
{trialG}), so one has to decompose $\Psi (x,0)$ using Fourier series: 
\begin{equation}
\Psi (x,0)=\frac 1{2\cdot \pi }\cdot \int_{-\infty }^\infty \widetilde{\psi }%
(k)\cdot e^{-i\cdot k\cdot x}dk,  \label{fourier}
\end{equation}
next find the initial conditions corresponding to the general component $%
\widetilde{\psi }(k)\cdot e^{-i\cdot k\cdot x}$, then one has to solve the
differential equation for $\eta (t,k)$ and $\mu (t,k),$ and finally compute
the inverse Fourier Transform.

Here I present an algorithm which can be used in solving the Schr\H{o}dinger
equation for a general form of the time and position dependence of potential
energy, without referring to the Fourier Transform. One can test Schr\H{o}%
dinger equation using the following explicit Euler substitution: 
\begin{equation}
\Psi (x,t)=N\cdot exp[\sum\limits_{n=0}^\infty \alpha _n(t)\cdot x^n]
\label{trialS}
\end{equation}
I used the series expansion of the potential energy w.r.t. $x$: 
\begin{equation}
V(x,t)=\sum\limits_{n=0}^\infty \frac 1{n!}V_x^{^{\prime }(n)}(0,t)\cdot x^n
\label{poten}
\end{equation}
The initial conditions are given by: 
\begin{equation}
\Psi (x,0)=N\cdot exp[\sum\limits_{n=0}^\infty \alpha _n(0)\cdot x^n]
\label{init}
\end{equation}
Expanding in power series, Schr\H{o}dinger equation gives: 
\begin{equation}
\dot{\alpha}_n(t)-\frac{i\cdot \hbar }{2\cdot m}[(n+2)\cdot (n+1)\cdot
\alpha _{n+2}+  \label{recur}
\end{equation}
\[
+\sum\limits_{k=0}^n(k+1)\cdot (n-k+1)\cdot \alpha _{k+1}\cdot \alpha
_{n-k+1}]+\frac 1{n!}V_x^{^{\prime }(n)}(0,t)=0
\]
One can easily note that if one has non-zero initial coefficients only for $%
n<n_0$ (which includes $\frac 1{n!}V_x^{^{\prime }(n)}(0,t)$), all
coefficients for $n<2\cdot n_0$ have to remain zero at every subsequent
moment (one can perform any-order time derivations in (\ref{recur}) at $t=0$
and obtain null values). Relation (\ref{recur}) can be very useful, either
for analytical calculations, or for numerical (finite difference) calculus
algorithms: 
\begin{equation}
\alpha _{n,p+1}=\alpha _{n,p}+\frac{i\cdot \hbar \cdot t_0}{2\cdot m}%
[(n+2)\cdot (n+1)\cdot \alpha _{n+2,p}+  \label{iter}
\end{equation}
\[
+\sum\limits_{k=0}^n(k+1)\cdot (n-k+1)\cdot \alpha _{k+1,p}\cdot \alpha
_{n-k+1,p}]+\frac 1{n!}V_x^{^{\prime }(n)}(0,p\cdot t_0)
\]
In (\ref{iter}) a finite time step $t_0$ was chosen . It is obvious that one
can iterate (\ref{iter}) in order to estimate $\{\alpha _{n,p}\}_n$ for
every moment $p\cdot t_0$ as functions of the initial coefficients $\{\alpha
_{n,0}\}_n$ given by (\ref{init}).

\narrowtext

\end{document}